# Object Oriented Information Computing over WWW

Dr. Pushpa R. Suri[1] and Harmunish Taneja[2]

[1]Department of Computer Science and Applications, Kurukshetra University
Kurukshetra- 136119, Haryana, India.

[2] Department of Information Technology, Maharishi Markendeshwar University,
Mullana, Haryana- 133203, India

**Abstract**
Traditional search engines on World Wide Web (WWW) focus essentially on relevance ranking at the page level. But this lead to missing innumerable structured information about real-world objects embedded in static Web pages and online Web databases. Page-level information retrieval (IR) can unfortunately lead to highly inaccurate relevance ranking in answering object-oriented queries. On the other hand, Object Oriented Information Computing (OOIC) is promising and greatly reduces the complexity of the system while improving reusability and manageability. The most distinguishing requirement of today's complex heterogeneous systems is the need of the computing system to instantly adapt to vigorously changing conditions. OOIC allows reflecting the dynamic characteristics of the applications by instantiating objects dynamically. In this paper, major challenges of OOIC as well as its rudiments are recapped. The review includes the insight to PopRank Model and comparison analysis of conventional page rank based IR with OOIC.

## 1. Introduction

Traditional Information Retrieval (IR) broadly extracts and manages the relevant information based upon user's request. It assumes pages as the retrieval units and the content of a page as reliable. OOIC is currently of considerable interest in the information Computing field for Information Retrieval and Management (IRM). Also, the reliability hypothesis is no longer valid in the object retrieval context when multiple copies of information about the same object typically exist. These copies may be conflicting because of diversity of Web site qualities and the limited performance of current information extraction techniques. Web data management demands extracting and integrating object information from the Web. Existing Web search engines generally treat a whole Web page as the unit for retrieval and consuming. However, there are various kinds of objects embedded in the static Web pages or Web databases. Typical objects are products, people, papers, organizations, etc. If these objects can be extracted and integrated from the Web, powerful object-level search engines can be built to meet users' exponentially growing information needs more precisely. Object identification on the Web [4] has been developed in recent years. This technique made it possible to extract and integrate all the related Web information about the same object together as an information unit. The Web information unit(s) is called object(s). Different objects form an object graph based on the correlations among them. In the Web graph, different pages have different attractiveness according to their in-links. Technologies such as PageRank [10] and HITS [7] have been successfully applied to distinguish the popularity of different Web pages through analyzing the link structure in the Web graph. In order to help users quickly locate their interested objects, the popularity of collected objects is calculated. More popular the objects are, more likely the user will be interested in them. The distinctive quality of the object graph is the heterogeneity of links. As the objects are related to each other through different types of relationships, therefore the popularity of Web objects is computed by applying link analysis techniques. The conventional link analysis methods presume that all the links are with the same 'support" semantics and equally important, directly applying these methods would result in unreasonable popularity ranking. Rest of the paper is as follows. Section 2 discusses challenges for Object Oriented Information Computing (OOIC). Section 3 summarizes existing object based computing on web. Section 4 gives the rudiments of OOIC and Section 5 presents the PopRank model and finally article is concluded in Section 6.

## 2. Challenges in OOIC.

OOIC provides a reusable object oriented (OO) framework for use with object oriented programming systems that provides an IR shell that permits a structure to define an index class that includes word index objects and provides an extensible IR system that evaluates a query by comparing information contained in the user query with information contained in the word index objects that relates to stocked pages. Preprocessing operations on pages produces the information in word index objects and the pages relevant to the user query will be identified, thereby providing a query result. The IR system user can pack pages into the computer system storage, index pages so their information can be subject to a query search, and request query evaluation to identify and retrieve pages most closely related to the subject matter of a user query. But convenience and efficiency share the package with spectrum of open problems [14]:

- Heterogeneous Web sources. Objects of the same type are distributed in diverse Web sources, whose structure is highly heterogeneous.
- Popularity propagation factor assignment to links. Assigning popularity propagation factor for links of different types of relationships to make the popularity ranking reasonable is extremely challenging.
- Automatic learning. Approach to automatically learn the popularity propagation factors for different types of links using the partial ranking of the objects given by domain experts is required.
- Dynamic query handling. Dynamic queries are equivalent to the direct manipulation interaction style, combining rapid feedback and visual controls to "carve" searches on the fly.
- Faceted metadata search. Faceted metadata search is the ability to query simultaneously along multiple attributes or dimensions and is complex.
- Collaborative filtering. Collaborative filtering uses the input (ratings) of other users to offer suggestions to another user — frequently used for recommending text, products or movies.
- Multilingual searches. Multilingual searches involve pages in different languages, including translation services as appropriate.
- Visual searches. Visual searches represent query parameters in a non textual format where applicable, for example calendar displays.

## 3. Related Work

Object identification on the Web [4] has been developed in recent years. PopRank [14], is a method which considers both the Web popularity of an object and the object relationships for OOIC to compute the popularity score of the Web object. PopRank extends the PageRank model by adding a popularity propagation factor (PPF) to each link pointing to an object, and uses different propagation factors for links of different types of relationships. Large combinations of feasible factors are required to get a reasonable assignment of the propagation factor. The existing Web IR techniques cannot provide satisfactory solution to the Web object extraction task highly heterogeneous diverse Web sources. Wrapper deduction [2, 8], Web database schema matching [6, 12] made it possible to extract and integrate all the related Web information about the same object together as an information unit. PageRank technique [10] calculates the importance of a Web page based on the scores of the pages pointing to the page. Hence, importance of Web pages pointed by many high quality pages rises. PageRank and HITS algorithms [7] are shown to be special cases of the unified link analysis framework but all the links have the same authority propagation factors in the PageRank model; it could not be explicitly applied to object-level ranking problem and XML elements. XRANK [15] rank XML elements using the link structure of the database. Object Rank system [1, 3] applies the random walk model to keyword search in databases modelled as labelled graphs. A unified link analysis framework [13] called "link fusion" considers both inter and intra type link structure among multi type inters related data objects for searching.

## 4. Rudiments of OOIC

Major OOIC elements are composed of object, its attributes, block and element. Object is generally represented by a set of attributes A= {a1, a2, ……., am }. Objects create an environment for developing and deploying object-oriented web applications. Object over web can be defined as the principle data unit about which takes IR to object extraction and management. Applications created with such Objects can be deployed as web sites and other web based applications. The time and cost benefits of instant, object-oriented development fascinates e-commerce. Object Attributes $A = \{$ attr 1, attr 2, ..., attr n$\}$, are properties which describe the concept represented as objects. Different types of objects are used to represent the information for different concepts. OOIC assume the same type of objects follows a common relational schema: $R$(attr1, attr 2, ..., attr n)., and key attributes, $AK = \{$ attr $K$1, attr $K$2, ..., attr $nK\}$ $A$, are properties which can uniquely identify an object. An object block can be defined as a collection of information within a Web page that relates to a single object. With the help of data record mining techniques, the object blocks from a page over WWW can be detected. The object block found on a Web page, can be further segmented to atomic extraction entities called object elements.

## 5. OOIC: THE POPRANK MODEL

PopRank model [14] is used for OOIC of the popularity of the different types of Web objects O1, O2, . . . ,O$n$ in an application domain. Objects are contained over WWW pages or the repositories, so the popularity of these pages and databases could also affect the popularity of their contained objects. If a page is popular, its object information is more likely to be read. Web popularity is used to denote the probability that a Web user reads the information about an object. OOIC can take place either via random surfing or through some Web page search engines. The object information is usually represented as a block of a Web page [9, 14]. The Web popularity can be computed by considering the PageRank scores of pages containing the object and the importance of the page blocks [5, 11]. PopRank model assumes that the Web records will uniformly proliferate its popularity (i.e. PageRank) to their objects, so the Web popularity from these sources can be easily computed. The popularity rank of an object [14] is calculated through the relationship graph of the objects in a domain.

5.1 Page Level V/S Object Level Search

In the PopRank model, random object finder model is used to explain the user's behaviour. This model simply keeps clicking on successive Web page links on WWW, page to object links, and object relationship links at random. The major differences between object level search and the page level search is summarized in table 1. In the page level search, information is extracted on the basis of the keywords entered by the user.

Table 1: From Page level search to Object level search.

|  | **Page-Level Search** | **Object-Level Search** |
|---|---|---|
| **Technology** | Information Retrieval Pages as Retrieval Units | Objects as Retrieval Units Database Machine Learning |
| **Data handling** | Page level | Object-level ranking and mining |
| **Model** | PageRank | PopRank |
| **Search** | Less accurate | More accurate |
| **Mining** | Conventional | Intelligent |
| **Repository** | Web databases | Object warehouses |
| **Advantages** | Ease of Use Ease of Authoring | Powerful Query Capability Aggregate and direct Answer |
| **Disadvantages** | Inadequate Query Capability | Object Identification is not trivial |

The outcome in this case is less precise because whole of the web database is used for the purpose. In case of Object level search, the information is extracted from a domain resulting accurate outcome. Moreover intelligent extortion of information is there due to domain specific search as object warehouses are used for computing the information.

For example, a research scholar may, use the search engines like altavista, google, alltheweb, excite or CiteSeer etc. to find several seed objects which could be some related papers, conferences/ journals, workshops, seminars or authors. After that she/he most likely just pursues the object relationship links to locate more manuscripts. She/he may want to extract the manuscripts cited by the manuscripts already read, or read papers of her/his favourite authors, journals/workshop/conferences. A manuscript cited by a large number of popular manuscripts could be popular, and a recent manuscript published in a reputed journal with few citations could also be popular. Scholar starts her/his random walk on the Web, and will start following the object relationship links once the first object on the Web is found, never hitting back but sooner or later gets bored and will restart his random walk on the Web again to find another seed object. In vertical search engine for OOIC, crawler fetches data over WWW associated with the targeted objects within a precise domain, and the crawled data is classified into diverse categories. A specific entity extractor is built to extract objects from the web data for each class. At the same time, same object information is cumulated from varied data sources. Once objects are extracted and aggregated, they are put into the object warehouses, and vertical search engines can be constructed based on the object warehouses. Moreover, advanced object-level ranking and mining techniques can be applied to make search more accurate and intelligent.

## 6. Conclusions

OOIC collect Web information for objects relevant for a specific application domain and rank objects in terms of their relevance and attractiveness to respond user queries. Also the analysis as shown in table 1 justifies that customary PageRank model is no longer valid for object popularity calculation major because of the existence of heterogeneous relationships between objects. This paper review *PopRank*, a domain-independent object-level link analysis model to rank the objects within a specific domain. OOIC assign a PPF to each type of object relationship; study how different PPFs for the heterogeneous relationships could affect the popularity ranking.

**Dr. Pushpa R. Suri** received her Ph.D. Degree from Kurukshetra University, Kurukshetra. She is working as Associate Professor in the Department of Computer Science and Applications at Kurukshetra University, Kurukshetra, Haryana, India. She has many publications in International and National Journals and Conferences. Her teaching and research activities include Discrete Mathematical Structure, Data Structure, Information Computing and Database Systems.

**Harmunish Taneja** received his M.Phil. degree in (Computer Science) from Algappa University, Tamil Nadu and Master of Computer Applications from Guru Jambeshwar University of Science and Technology, Hissar, Haryana, India. Presently he is working as Assistant Professor in Information Technology Department of M.M. University, Mullana, Haryana, India. He is pursuing Ph.D. (Computer Science) from Kurukshetra University, Kurukshetra. He has published 11 papers in International / National Conferences and Seminars. His teaching and research areas include Database systems, Web Information Retrieval, and Object Oriented Information Computing.